\documentstyle[epsfig,prd,aps]{revtex}
\begin{document}
\draft
%
%
\twocolumn[\hsize\textwidth\columnwidth\hsize\csname
@twocolumnfalse\endcsname
\preprint{astro-ph/0104112}
\title{The Tensor to Scalar Ratio of Phantom Dark Energy Models}
\author{A. E. Schulz}
\address{Department of Physics, Harvard University,
Cambridge, Massachusetts 02138}
\author{Martin White}
\address{Harvard-Smithsonian CfA, 60 Garden St, Cambridge, MA 02138}
\vspace{3 mm}
\date{\today}
\maketitle
\begin{abstract}

We investigate the anisotropies in the cosmic microwave background in a class
of models which possess a positive cosmic energy density but negative pressure,
with a constant equation of state $w = (p/\rho) <-1$.
We calculate the temperature and polarization anisotropy spectra for both
scalar and tensor perturbations by modifying the publicly available code
CMBfast.
For a constant initial curvature perturbation or tensor normalization, we have
calculated the final anisotropy spectra as a function of the dark energy
density and equation of state $w$ and of the scalar and tensor spectral
indices.  This allows us to calculate the dependence of the tensor-to-scalar
ratio on $w$ in a model with phantom dark energy, which may be important for
interpreting any future detection of long-wavelength gravitational waves.
\end{abstract}

\vskip2pc]

\section{Introduction}

For many years \cite{EARLY}, arguments have been made in favor of the
$\Lambda$CDM model because of its desirable property of having spatial
flatness while accommodating a low matter density; conditions which seem
to be favored by current astrophysical observations.
A few years ago, however, an old alternative
\cite{xmat}
resurfaced and has been receiving considerable attention \cite{later}. This
alternative, which we will refer to as $\Phi$CDM, borrows from the 
literature on inflation and postulates a slowly rolling scalar field whose
energy density has only recently become cosmologically relevant.
This slowly rolling scalar field provides a constituent of the universe which
is smooth below the horizon scale\footnote{The speed of sound for a scalar
field is the speed of light.} and whose evolution can be made arbitrarily slow.
This is usually parameterized by the equation of state, $w\equiv p/\rho$,
with the model becoming indistinguishable from the cosmological constant in
the limit $w\to -1$.  Generically $1>w>-1$ and $w$ is typically a function
of time.

Of particular interest to us was a paper by Caldwell \cite{phantom} which
introduced a $\Phi$CDM model with $w<-1$.
This form of dark energy, dubbed {\sl phantom energy\/} also has the
remarkable property of satisfying $\rho_{\phi}+p_{\phi}<0$, 
just as in the standard
$\Phi$CDM case.
It was shown that this model is consistent with both recent observations and
classical tests of cosmology, in some cases providing a better fit than the
more familiar models with $w>-1$.

There are several known ways to achieve $w<-1$ e.g.~Ref.~\cite{ParAve},
but Caldwell's model makes use of a non-canonical kinetic term.
There is no obvious motivation from particle physics for considering such a
radical extension to the theory, and indeed the model has been criticized on
these grounds \cite{ParAve}.
However we regard this as a useful toy model because it allows us to simply
extend the equation of state parameter space continuously below $w=-1$ without
the additional complications inherent in modifications of the gravitational
sector of the theory.
(We shall discuss how our results depend on this detail later.)
Observational constraints are often maximally likely at the value $w=-1$ and
for the purposes of  maximum likelihood analysis it is better not to have the
region of maximum likelihood bumping against the edge of the parameter space.
Thus it is advantageous to extend the domain of the parameter $w$, even
if one is primarily interested, on theoretical grounds, in values $w\ge -1$.

One aspect which was not studied in \cite{phantom} was the effect of
gravitational wave perturbations in the CMB, and in particular the dependence
of the tensor-to-scalar ratio on the equation of state $w$.
In this paper we study the tensor anisotropy spectrum in the phantom model.
We show that the normalization of the tensor perturbations in
the CMB is only weakly dependent on the equation of state.
The behavior of the tensor-to-scalar ratio is therefore dominated by the
$w$ dependence of the scalar perturbations.

We are interested in the tensor contribution to the anisotropy spectrum
because it is, in principle, observable and directly connected to the energy
scale of inflation\cite{book}, a fundamental parameter of considerable
importance!
Our calibration of the relation between the amplitude of the anisotropy in
the CMB and this fundamental energy scale is clearly necessary to interpret
any future observations of this signal.

\section{The Model}
 
Phenomenologically, the property of negative pressure and positive energy
density can be achieved by considering the (unorthodox) Lagrangian density
\begin{equation}
{\cal L} =
  {1\over2} g^{\mu \nu}\partial _{\mu} \phi \partial _{\nu} \phi -V(\phi)
\end{equation}
for the phantom component $\phi$.  We are adopting the metric convention 
$g_{\mu\nu}={\rm diag}\left({-a^2, a^2, a^2, a^2}\right)$ where $a(t)$ is the
cosmological scale factor.
To lowest order the non-canonical negative kinetic term in the Lagrangian
produces the following expressions for the pressure and energy density
of this dark energy:
\begin{equation}
  \rho_\phi = -{1 \over 2a^2} \dot{\phi}^2 + V(\phi)
\end{equation}
\begin{equation}
  p_\phi = -{1 \over 2a^2} \dot{\phi}^2 - V(\phi)
\end{equation}
where over-dots represent derivatives with respect to conformal time
$\eta$, and $dt=a d \eta$.  Negating the kinetic term also alters the
equation of motion of the field from that of a field with a canonical
Lagrangian by switching the sign of the $\partial V/\partial \phi$ term.
\begin{equation}
  \ddot{\phi} + 2 {\dot{a} \over a} \dot{\phi} - {1\over a^2}
  {\partial V\over \partial \phi} = 0
\end{equation}
Clearly the evolution of the scalar field is coupled into the evolution of
the background Friedmann equations. 
\begin{equation}
  {\cal H}^2 =\left( {\dot a\over a}\right)^2 = 
{8\pi G\over 3}\ a^2 \sum_i \rho_i
\end{equation}
Here ${\cal H}=a H = a (a^{-1} da/dt)$, taking into account that our
derivatives are with respect to conformal time $\eta$.
The effect of $\phi$ on the evolution of the background cosmology can be
implemented with only a slight modification to the CMBfast integration code
\cite{cmbfast} to include the new component's pressure and energy density.

Once the expressions for $\rho_{\phi}$ and $p_{\phi}$ are obtained, it is
simple to invert them to obtain the potential, $V$ in terms of the constant
$w$ and derivatives of the scalar field.
In order to prevent tachyonic modes of the scalar field from developing,
we address only models with constant $w$ \cite{phantom}:
\begin{equation}
  V= -{{\dot{\phi} ^2} \over {2 a^2}} \left({{1-w} \over {1+w}}\right)
\end{equation}
Following \cite{phantom}, by treating the dark energy as a perfect fluid,
we can obtain an explicit expression for $\dot{\phi}$.
\begin{equation}
  \dot{\phi}=\sqrt{-(1+w)\rho_{\rm crit}\Omega_\phi}\quad a^{-(1+3w)/2}
\end{equation}
where $\rho_{\rm crit}=3{\cal H}^2/8\pi Ga^2$, and
$\Omega=\rho_{\phi}/\rho_{\rm crit}$.
Taking its derivatives we observe
\begin{equation}
  {\partial V \over \partial \phi}=
  (1-w){3{\cal H} \dot{\phi} \over 2 a^2}
\end{equation}
\begin{equation}
  {\partial^2 V \over \partial \phi^2}=
  (1-w){3\over 2 a^2} \left[ \dot{\cal H}-{\cal H} ^2\left({5\over 2}-
     {3w\over 2}\right) \right]
\end{equation}
Thus $V$ and its derivatives will be
functions of $a$ and the parameters $w$, $\Omega_{\phi},$ and $\rho_0$ only.

To calculate the anisotropy in the CMB we must include the perturbations of
$\phi$ to first order.  We work throughout in the synchronous gauge.
Since we are solving linearized equations it is useful to Fourier transform
the perturbations so that we can solve the differential equations for each
$k$-mode independently.
The equation of motion for a particular $k$-mode is
\begin{equation}
  \ddot{\delta\phi}_k+
  2{\dot{a}\over a}\dot{\delta\phi}_k+
  a^2\left(k^2-{\partial^2V\over\partial\phi^2}\right)\delta\phi_k
  = -{1\over2}\dot h\dot{\phi}
\label{eqn:evolve}
\end{equation}
where $h$ is the trace of the metric perturbation in the synchronous gauge.
The scalar field perturbation, being of spin-0, does not depend on the other
metric perturbation directly.
The $\delta\phi_k$s produce the fluctuations in the energy density and
pressure of the $\phi$ field:
\begin{equation}
  \delta\rho_{\phi}=\left({2\over1-w}\right)
  {\partial V \over\partial\phi}\delta\phi_k \qquad \phantom{.}
\end{equation}
\begin{equation}
  \delta p_{\phi}=\left({2 w\over1-w}\right)
  {\partial V \over\partial\phi}\delta\phi_k \qquad .
\end{equation}
These stress-energy perturbations must be included in the evolution of the
metric perturbations for a self-consistent solution.
The $\Phi$CDM component contributes a source term to the right hand side 
of the Einstein equations.  
To make these calculations, we have modified CMBfast by including the 
$\Phi$CDM contribution 
to the evolution of the background cosmology in the Friedmann 
equations.  We have also added the contribution of 
density and pressure perturbations 
of the $\phi$ field to the total curvature perturbation, using the results
derived above.  Even though the background pressure is negative, we are 
certain that these perturbations are stable, because $\partial^2 V/ \partial
\phi^2$ is a negative definite quantity in this case.  Examining 
Eq.~(\ref{eqn:evolve}) we see that the square of the 
effective mass will never become negative, and exponentially growing 
tachyonic modes will not occur.  

Note that $\delta\phi$ cannot source gravitational wave modes because $\phi$
is a scalar field, while a source of gravitational waves must have spin 2.
Therefore, the only effect of $\phi$ on the gravitational wave anisotropies
is through the change in the evolution of the scale factor \cite{tenscmb}.

\section{CMB Anisotropy Spectra}

\begin{figure}
\begin{center}
\leavevmode
\epsfxsize=3.25truein\epsffile{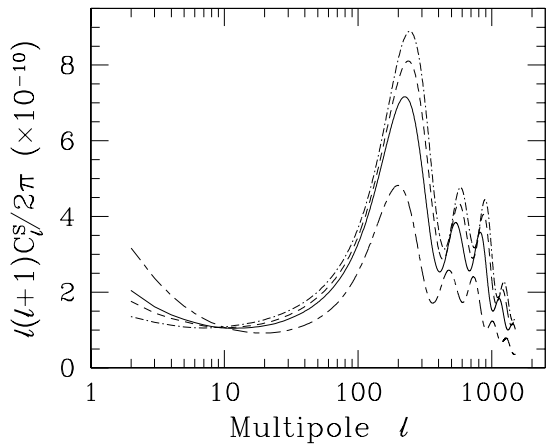}
\end{center}
\caption{The scalar CMB temperature
anisotropy spectra,
$\ell(\ell+1)C_{\ell}/2(\pi)$ vs.$\ell$ for several values of $w$.
The solid lines are $w=-1.1$, dashed are $w=-3$, dot-dash are $w=-10$,
and dash-dash are $w=-0.4$.  The spectra have been normalized
at $C_{10}$ to fit the {\sl COBE\/} data.}
\label{fig:scalar}
\end{figure}

\begin{figure}
\begin{center}
\leavevmode
\epsfxsize=3.25truein\epsffile{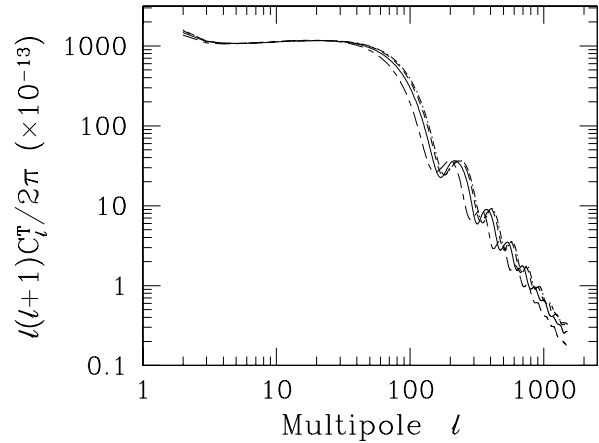}
\end{center}
\caption{The tensor CMB temperature 
anisotropy spectra as a function of
$w$.  As in Fig.~\protect\ref{fig:scalar}, the solid lines are $w=-1.1$,
dashed are $w=-3$, dot-dash are $w=-10$, and dash-dash are $w=-0.4$.
Notice that the curves are almost degenerate for the first 40 multipoles.
There appears to be negligible dependence of the tensor normalization
on the equation of state $w$.}
\label{fig:tensor}
\end{figure}

Figs~\ref{fig:scalar} and \ref{fig:tensor} display the scalar and tensor
components of the temperature anisotropy for several values of the equation
of state parameter $w$.
In each case we assumed that the cosmological parameters were
$h=.67$, $\Omega_m=0.3=1-\Omega_\phi$, $\Omega_b=0.04$ and that the
scalar spectral index $n_s=1$, and the tensor spectral index $n_t=0$.
It is standard practice \cite{ARAA} to write the temperature field on the
sky as a sum of spherical harmonics with coefficients $a_{\ell m}$.
The multipole moments, $C_{\ell}$, are then defined as
$C_{\ell}\equiv |a_{\ell m}|^2$
and the power spectrum, $\ell(\ell+1)C_\ell/(2\pi)$, is approximately the
power per logarithmic interval in $\ell$.

\begin{figure}
\begin{center}
\leavevmode
\epsfxsize=3.25truein\epsffile{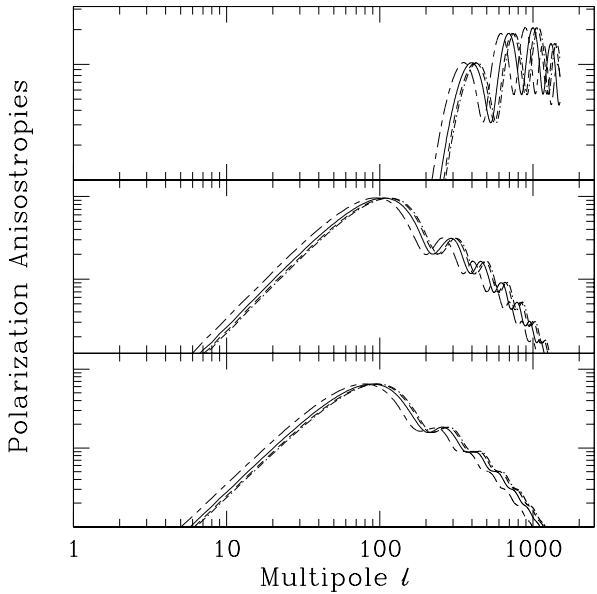}
\end{center}
\caption{The polarization anisotropy spectra.  For the scalar modes, the
E-E polarizations anisotropies are shown (top), and for the tensor modes
the E-E and B-B polarization anisotropies are presented (middle and bottom).
The scale is arbitrary, depending on our particular 
choice of normalization.}

\label{fig:polar}
\end{figure}

Since we are working to linear order in perturbation theory both the scalar
and tensor spectra have one free overall normalization (ignoring for a moment
any constraints imposed by e.g.~inflation).
In our calculations we fixed the scalar normalization by requiring that the
initial spatial curvature in the total matter gauge \cite{TMG} $\zeta=1$.
For the tensor normalization\footnote{The tensor perturbation is gauge
independent.} we have chosen $h=1$ initially.
This choice makes it easy to interprate the dependence on $w$, and also to
calculate the cosmology dependent fitting function presented in
\S\ref{sec:fits}.
However with such a normalization the curves are very similar beyond
$\ell\sim100$ for all values of $w$ so for plotting purposes we have
normalized the curves at $\ell=10$ in Figs~\ref{fig:scalar} and
\ref{fig:tensor}.
In Fig.~\ref{fig:polar}, the E-E and B-B (tensors only) polarization spectra
are presented with $\zeta=1$ or $h=1$ normalization, which illustrates clearly
that the height of the features is uniform with $w$.

Consider first  the anisotropy spectrum in Fig.~\ref{fig:scalar}.
There are three features exhibiting $w$ dependence:
the height of the first peak, the height of the tail at low multipoles,
and the location of the peaks.
The change in the height of the first peak traces to our decision to
re-normalize the spectra to fit the {\sl COBE\/} data.
In fact the first two effects are manifestations of the same physical
property: the suppression of the late Integrated Sachs-Wolfe (ISW) 
effect for lower $w$.
In models such as $\Phi$CDM, a period of accelerated expansion begins
in the universe when the scalar field $\phi$ becomes the dominant 
form of energy density.  For a fixed matter density $\Omega_m$ (measured
relative to the critical density) 
the value of the parameter $w$ determines how early matter-$\phi$
equality occurs, and therefore how long the universe has been
experiencing the current period of accelerated expansion.
When an accelerated expansion is occurring, potential wells
which had formed earlier as a result of gravitational collapse begin 
to decay.   If the well is decaying, photons which fall into the well 
gain more energy than they lose climbing out, which generates an 
anisotropy.  This is the dominant effect at low $\ell$. 
For $w<-1$ models, equality happens even later than in the case of the
cosmological constant, while for $w>-1$, it happens earlier.
For very negative values of $w$, there has been almost no acceleration in the
recent history, so the potential wells have not decayed as dramatically
as they do in $\Lambda$CDM models.
As a result, the late Integrated Sachs-Wolfe (ISW) effect has been almost
entirely turned off \cite{phantom}.

When normalized to $\zeta=1$, the effect on the spectrum is to enhance the
amplitude at low multipoles as $w$ decreases.
Re-normalizing to fit the {\sl COBE\/} data at $\ell=10$, the enhanced
low-$\ell$ tail translates into a change in the peak height.
For energy density perturbations, the dependence of the spectrum normalization
on the equation of state is principally due to the lack of late ISW effect
as $w$ decreases.

The normalization of the tensor spectrum appears not to be affected by $w$
in any significant way (Fig.~\ref{fig:tensor}).
This can be seen by noting that the peaks are all roughly at the same height,
and the spectra between $2<\ell<10$ agree closely
(compare with Fig.~\ref{fig:scalar}).
The small $w$ dependence that can be seen occurs for the lowest multipoles,
corresponding to largest angular scales, and is again due to the ISW effect
on the tensor spectrum at late times.

For all the anisotropy spectra (both 
temperature and polarization anisotropies)
there is a universal 
shift to the right for lower values of $w$.
The reason for the shift is the same in all cases.
The $\ell$ position of any feature in a spectrum depends upon the
(comoving) angular diameter distance to the last scattering surface.
This distance depends on $w$ as follows:
\begin{equation}
  d_{\rm lss}  =\int_{a_{\rm lss}}^1 {da\over a^2 H}
\end{equation}
where
\begin{equation}
  H^2(a) = \Omega_r a^{-4} + \Omega_m a^{-3} + \Omega_{\phi} a^{-3(1+w)}
\end{equation}
One notes that if $w$ is more negative, this integral is larger, which 
implies that the last scattering surface is further away.  A feature of a
given physical size thus subtends a smaller angle, shifting the anisotropy
spectra to higher values of $\ell$.  In the case of the polarization 
anisotropies, this is the only dependence on $w$ (Fig. ~\ref{fig:polar}). 

\section{dependence of $T/S$ on \lowercase{$w$} and $\lowercase{n}_S$}
\label{sec:fits}

For $\Omega_b = .04$ and assuming a flat universe, we have explicitly
calculated the evolution of the normalization of the spectrum from an initial
curvature perturbation $\zeta=1$ and from $h=1$.
The normalization at the tenth multipole is equal to the value just after
primordial inflation, multiplied by a cosmology dependent
``transfer function'', which we will call $f_S$ and $f_T$.
Using the method described in \cite{turnwhiteII} we have numerically
determined this function for the ranges $-2.5<w<-0.5$,
$0.6<\Omega_{\phi}<0.8$, and $-0.3<n_T<0.3$ and provide fitting functions
below.
The tensor spectral index $n_T$, is related to the scalar 
spectral index $n_T=n_S-1$ for power law models of primordial 
inflation.
We have  deliberately left a $w$ term in the evolution of the tensor
normalization to illustrate how insignificant it it compared to the terms
proportional to the spectral index and $\Omega_{\phi}$.
A fit is presented below, where $a=0.6156$,$b=1.6$,and $c=-.2$. The scalar 
transfer function has a mean accuracy of $2\%$ over the range indicated above,
and at worst is off by $10\%$ near the edges of the parameter domains.  This
would correspond to a worst case $5\%$ error in the temperature. Since the 
fit begins to go bad near the edges of the parameter space, we do not 
recommend extrapolating these functions further.  
The tensor function has a mean accuracy of 
$.8\%$, with a worst case of $6\%$ for a few 
points at the very edge of the fit.
\begin{eqnarray}
  f_S(w,n_T,\Omega_{\phi}) 
=-1.341+6.031 \Omega^a n^b (-w)^c +1.936 \Omega^a \nonumber\\+ 3.312 n^b
+4.584 (-w)^c -5.148 \Omega^a n^b -9.375 \Omega^a (-w)^c \nonumber\\
-1.946 n^b (-w)^c +5.172 \Omega (-w)^c -2.692 n (-w)^c 
\nonumber \\-.072 n^b w
-.211 (n-1)^2
\end{eqnarray}
\begin{eqnarray}
  f_T(w,n_T,\Omega_{\phi})= 0.358 + 0.001 w  +0.822n_T 
-0.046\Omega_{\phi} \nonumber \\
  + 0.337 n_T^2 
  +0.029 \Omega_{\phi}^2 - 0.487 n_t\Omega_{\phi}
  \qquad .
\end{eqnarray}
The tensor to scalar ratio is given by
\begin{equation}
  {C^T_{10} \over C^S_{10}} =
  {f_T(w,n_T,\Omega_{\phi})\over f_S(w,n_T,\Omega_{\phi})}
  \ {A_T^2\over A_S^2}
\label{eqn:tovers}
\end{equation}

\begin{figure}
\begin{center}
\leavevmode
\epsfxsize=3.25truein\epsffile{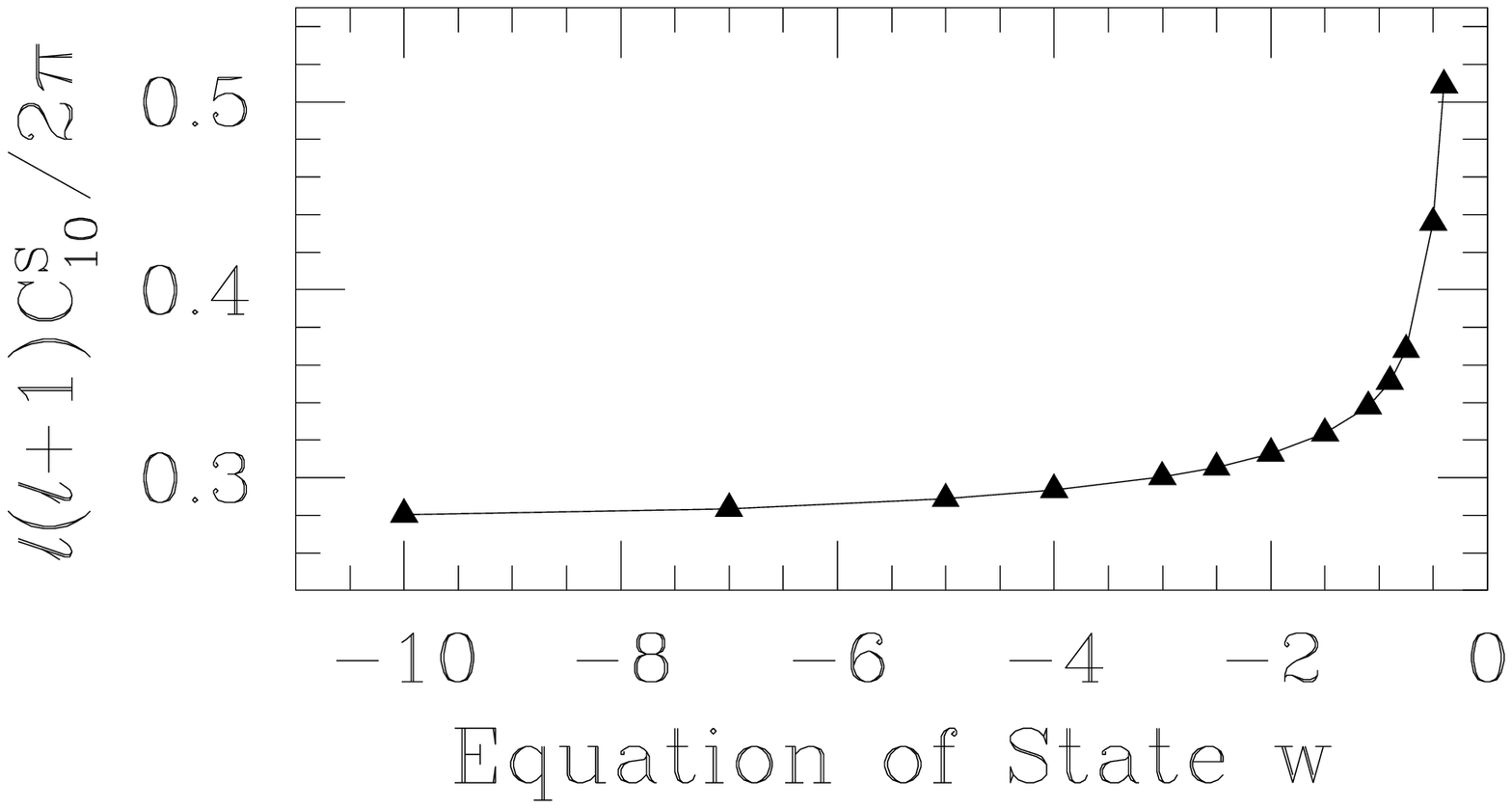}
\epsfxsize=3.25truein\epsffile{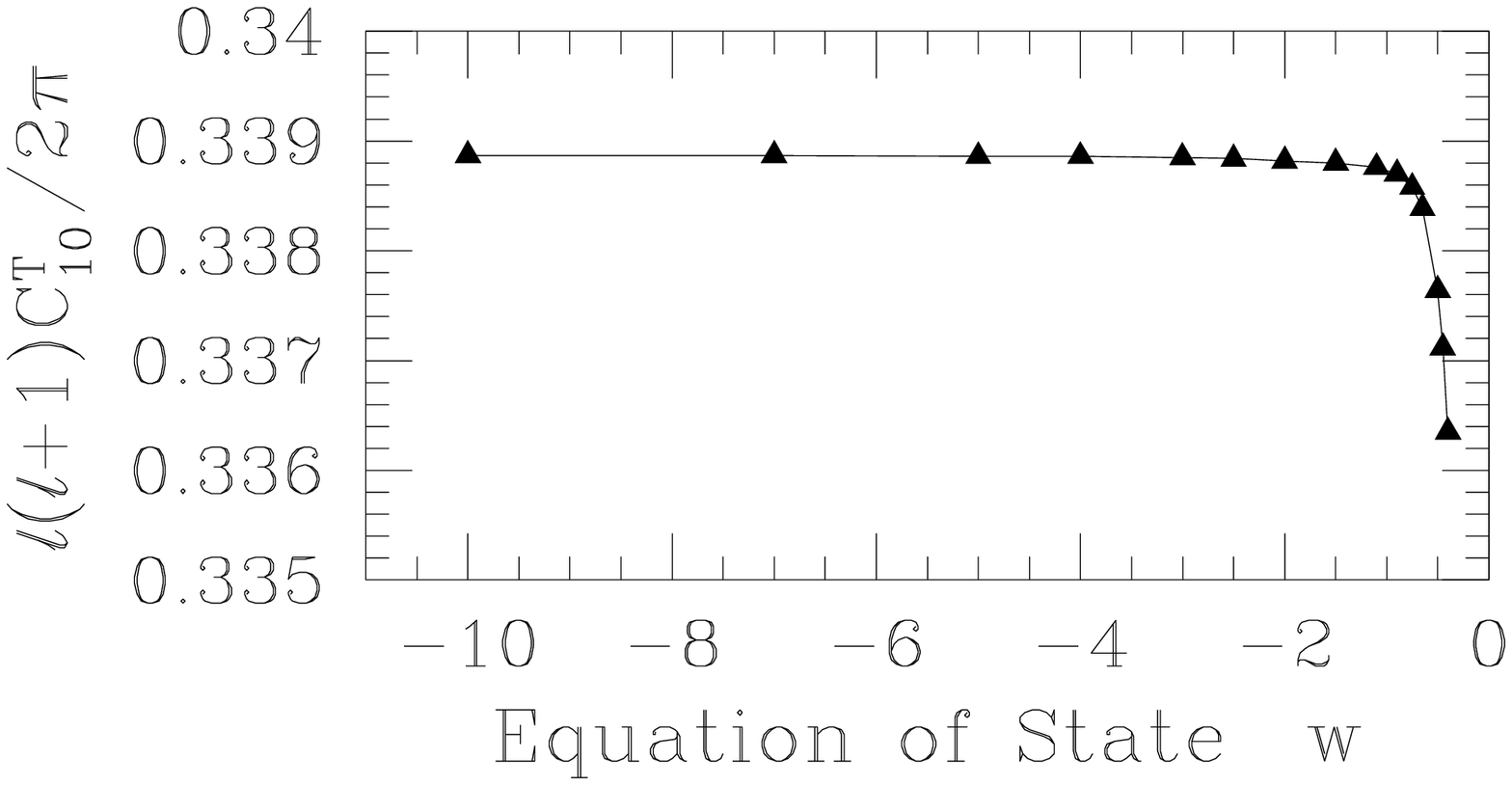}
\end{center}
\caption{The normalization of the tensor and scalar spectra at the 10th
multipole for an initial perturbation $\zeta=1$ and $h=1$.}
\label{fig:c10}
\end{figure}

Fig.~\ref{fig:c10} shows the dependence of both the scalar and tensor
normalizations at $C_{10}$ on the equation of state $w$ for an even broader
domain than was used in the fitting formulae above.
Observe that even over this huge range in the equation of state, the tensor
normalization changes by less than $1\%$.

\section{Discussion}

We have extended the work of \cite{phantom} on the $\Phi$CDM model by
calculating the tensor component of the CMB anisotropies.
We have shown that the normalization of the tensor component is essentially
independent of the equation of state $w$ in this model.
For a fixed value of the initial curvature perturbation after inflation
$\zeta=1$ and a fixed initial gravity-wave amplitude, we have quantified
the $w$, $\Omega_{\phi}$ and $n_T$ dependence of both the tensor and scalar
anisotropy normalizations at $C_{10}$ for a broad domain of $w$.
We displayed the anisotropy spectra for several values 
of $w$ and discussed the features.  

Most of the results we present for $w<-1$ are only weakly sensitive to the
particular model chosen \cite{phantom}.
While truly general statements are not possible, we can make several points
about model dependence.
The fact that our acceleration is driven by a field of spin zero implies there
is no source of gravitational waves as $\phi$ begins to dominate.
This would remain true for other scalar (or vector) driven theories, while
more complicated assumptions (e.g.~higher order gravity theories) may have
explicit source terms.
However, we would expect only the longest wavelength gravitational waves to
differ due to this in any significant way.
This is because the acceleration takes place very late in the history of the
universe for $w<-1$, so only the longest modes are affected.
The longest wavelength modes contribute primarily to the very low-$\ell$
moments of the spectrum.
A similar argument can be made for the scalar modes, where the overall growth
rate should be independent of the detailed model, but the longest wavelength
modes will depend on any fluctuations in the ``phantom'' component through
their impact on the evolution of the metric perturbations.
Finally, the angular shift with $w$ of features in the spectrum is robust.

We have quantified the dependence of the tensor to scalar ratio
${C^T_{10}/C^S_{10}}$ on $w$, $\Omega_{\phi}$, and $n_T$.
For the model of Ref.~\cite{phantom}, Eq.~(\ref{eqn:tovers}) allows us to
relate the (potentially) observable ${C^T_{10}/C^S_{10}}$ to the energy scale
of inflation, in the event that a gravitational wave signal is actually
observed from the CMB.

\section*{Acknowledgments}

Many thanks to J. D.~Cohn for countless useful discussions on this work. 
We thank Douglas Scott for comments on the manuscript, Mattias Zaldarriaga
for providing a hitherto unreleased version of the CMBfast code, and also
Max Tegmark for his advice on resolving various gauge confusions.

This work was supported in part by the Alfred P. Sloan Foundation and the
National Science Foundation, through grant PHY-0096151.

\end{document}